\documentclass{article}%
\usepackage{amssymb}
\usepackage{amsmath}
\usepackage{amsfonts}
\usepackage{graphicx}%
\usepackage{epsfig}
\begin{document}
 \title{STREAMING CAUSED  BY  IMPULSE  ACOUSTIC  WAVE. }
\author{  Anna Perelomova
\small\\  Gdansk University of Technology, \\
ul Narutowicza 11/12, 80-952 Gdansk, Poland,
\small \\ anpe@mifgate.pg.gda.pl \\  \\[2ex] }
\maketitle

\renewcommand{\abstractname}{\small Abstract}
 \begin{abstract}
 Five
eigenvectors of the linear thermoviscous flow over the homogeneous
background derived for the quasi-plane geometry of the flow. The
corresponding projectors are calculated  and applied to get the
nonlinear evolution equations for the interacting vortical and
acoustic modes. Equation on streaming cased by arbitrary acoustic
wave is specified. The correspondence to the known results on
streaming cased by quasi-periodic source is traced. The radiation
acoustic force is calculated for the mono-polar source.
\end{abstract}

\section{INTRODUCTION}

Traditionally, streaming is referred as a localized mean flow caused by an
acoustic source. This flow is created by the Reynolds forces, non-zero
time-averaged values of the quadratic acoustic terms arising in the equation
of conservation of momentum. There exist extensive reviews on this subject [1-3].

Steady acoustic streaming is described by the mean Eulerian velocity
$\langle\overrightarrow{v}\rangle$, or the mean mass flow velocity
$\overrightarrow{U}$, where $\overrightarrow{U}$ is defined by
$\overrightarrow{U}=\frac{\left\langle \rho\overrightarrow{v}\right\rangle
}{\rho_{0}}=\overrightarrow{V}+\frac{\left\langle \rho_{a}\overrightarrow
{v}_{a}\right\rangle }{\rho_{0}}.$Here, $\overrightarrow{V}$ is a slow
component of velocity, namely the Eulerian velocity of the streaming, index
$a$ labels acoustic wave, $\rho_{0}$ is undisturbed density. Square brackets
designate averaging over time whose interval is much shorter than the
transient time of streaming and much longer than a period of a sound wave. It
is pointed, that the difference between $\overrightarrow{U}$ and
$\overrightarrow{V}$ is roughly proportional to the acoustic intensity vector
and is small compared with any of mean velocities [4]. From the continuity
equation it follows $\nabla\overrightarrow{U}=0$ if we assume $\left\langle
\frac{\partial\rho}{\partial t}\right\rangle =0$.

The recent discussion of acoustic streaming [5,6] suggest that there is
important unresolved issue concerning acoustic streaming, namely the effect of
fluid incompressibility. The inconsistency of this point is that fluid is
supposed to be incompressible while the acoustic wave casing streaming may
propagate only over compressible medium. The next, the conservation of energy
usually is not considered. It is reasonable since excluding of one variable
(density) reduces an initial system. It has been proved that the effect of the
heat conductivity could not be discarded in a study of temperature variation
associated with the streaming [6]. Actually, this approximation is well
understood in a typical liquid like water but should be revised for other
liquids [7]. That is a reason to start from the full system of conservation
laws including the energy balance equation and the continuity equation.

The procedure of the temporal averaging needs a continuous acoustic wave of
many periods as a source, thus the well-known results on acoustic streaming
apply only for this type of acoustic source though quite realistic. There are
clear inconsistency concerning the very procedure of the temporal averaging.
As it has been mentioned before, an equality $\nabla\overrightarrow{U}=0$ is
correct only if $\left\langle \frac{\partial\rho}{\partial t}\right\rangle =0$
$.$ Actually, \ the overall field includes not only acoustic and vortical
modes, but also of the entropy mode. The entropy mode caused by acoustic wave
is known as a slow decrease of density and\ isobaric increase of temperature
of the background of acoustic wave propagation. Recently, the importance of
change in density of surrounding (not only increase in temperature) was
underlined in the review [8]. If nonlinear interactions are accounted in a
proper way, for density of the entropy mode one gets an equation :

$\frac{\partial\rho_{e}}{\partial t}+\frac{\eta}{\rho_{0}}\Delta\rho_{e}%
=S_{a}$,

where $\eta$ is shear viscosity, $S_{a}$ is an acoustic source proportional to
a gradient of acoustic intensity, index $e$ concerns the entropy mode. An
attenuation of acoustic wave leads to slow but continuous decrease of density
so $\langle\frac{\partial\rho_{e}}{\partial t}\rangle=0$ becomes incorrect as
well as $\left\langle \frac{\partial\rho}{\partial t}\right\rangle =0,$ since
$\rho=\rho_{0}+\rho_{a}+\rho_{e}.$ Therefore, if the acoustic heating caused
by periodic acoustic wave is accounted, $\nabla\overrightarrow{U}=0$ becomes incorrect.

These inconsistencies call new ways to theoretical investigation
of acoustic streaming, including streaming caused by non-periodic
acoustic wave. We plan to start with all conservation equations
for compressible thermoviscous flow, then to define modes by
establishing of the relations between perturbations (section 2),
to demonstrate how the well-known linear and nonlinear evolution
equations may be obtained by projecting (sections 3,4), and,
finally, to apply projectors to calculating of the streaming
caused by pulse acoustic wave (section 5).

The idea to decompose the linear flow into specific modes is not novel and has
been exploited for a long time , see [9] and referred papers, where
homogeneous background with sources of heat and mass are considered. The
concrete ideas of doing it automatically by using of projectors in the wide
variety of problems (also flow over inhomogeneous media like bubbly liquid
[10], flows affected by external forces including the gravitational one which
changes the background density and pressure [11], both one- and
multi-dimensional problems, see [12] ). The principle advance is an expansion
of the ideas into area of nonlinear flow: to get nonlinear coupled evolution
equations for interacting modes and to solve the system approximately. The
deriving of the final nonlinear system is algorithmic, one just acts by
projectors at the initial system of nonlinear equations. Some ideas concerning
separation and interaction of acoustic-gravity and Rossby waves may be found
in [13].

\bigskip

\section{ Modes and projectors for the flow with equations of
state in the general form.}

The mass, momentum and energy conservation equations for the thermoviscous
flow read:

$$\frac{\partial\rho}{\partial t}+\vec{\nabla}(\rho\vec{v})=0$$

$$\rho\left[  \frac{\partial\vec{v} }{\partial t} +(\vec{v}
\vec{\nabla} )\vec{v} \right]  =-\vec{\nabla} p+\eta\Delta\vec{v}
+\left( \varsigma +\frac{\eta}{3} \right)  \vec{\nabla}
(\vec{\nabla} \vec{v} )$$

$$\rho\left[  \frac{\partial e}{\partial
t}+(\vec{v}\vec{\nabla})e\right] +p\vec{\nabla}\vec{v}-\chi\Delta
T=\varsigma\left(  \vec{\nabla}\vec
{v}\right)  ^{2}+\frac{\eta}{2}\left(  \frac{\partial v_{i}}{\partial x_{k}%
}+\frac{\partial v_{k}}{\partial
x_{i}}-\frac{2}{3}\delta_{ik}\frac{\partial v_{l}}{\partial
x_{l}}\right)  ^{2} \eqno(1) $$

Here, among already mentioned variables, $p$ is pressure, $e,T$- internal
energy per unit mass and temperature; $\varsigma,$ $\chi$are bulk viscosity
and thermal conductivity coefficient (both supposed to be constants as well as
$\eta$ ), $x_{i}$ - space coordinates. Except of the dynamic equations (1),
the two thermodynamic relations are necessary: $e(p,\rho),T(p,\rho)$ . To
treat a wide variety of substances, let us use the most general form of these
relations as expansion in the Fourier series:

$$\rho_{0}e^{\prime}=E_{1}p^{\prime}+\frac{E_{2}p_{0}}{\rho_{0}}\rho^{\prime
}+\frac{E_{3}}{p_{0}}p^{\prime2}+\frac{E_{4}p_{0}}{\rho_{0}^{2}}\rho^{\prime
2}+\frac{E_{5}}{\rho_{0}}p^{\prime}\rho^{\prime}+\ldots$$

$$T^{\prime}=\frac{\Theta_{1}}{\rho_{0}C_{v}}p^{\prime}+\frac{\Theta_{2}p_{0}%
}{\rho_{0}^{2}C_{v}}\rho^{\prime}+\ldots \eqno(2) $$

The background values are marked by zero, perturbations are primed, $C_{V}$
means specific heat per unit mass at constant volume, $E_{1},...\Theta
_{1},...$are dimensionless coefficients,

$\Theta_{1}=\frac{\rho_{0}C_{v}k}{\beta},$ \ \ $\Theta_{2}=-\frac{\rho
_{0}C_{v}\beta}{p_{0}},$ \ \ \ \ \ \ \ \ \ \ \ \ \ \ \ \ \ \ \ \ \ \ \ \ \ \ \ \ \ \ \ \ \ \ \ \ \ \ \ \ \ \ \ \ \ \ \ \ \ \ \ \ \ \ \ \ \ \ \ \ \ \ \ \ \ \ \ \ \ \ \ \ \ \ \ \ \ \ \ \ \ \ \ \ \ \ \ \ \ \ \ \ \ \ \ \ \ \ \ \ \ \ \ \ \ \ \ \ \ \ \ \ \ \ \ \ \ \ \ \ \ \ \ \ \ \ \ \ \ \ \ \ \ \ \ \ \ \ \ \ \ \ \ \ (3)

where $k=\frac{1}{\rho_{0}}\left(  \frac{\partial p}{\partial\rho}\right)
_{T=T(p_{0},\rho_{0})},\beta=-\frac{1}{\rho_{0}}\left(  \frac{\partial\rho
}{\partial T}\right)  _{\rho=\rho_{0}}.$

For the tree-dimensional flow $\vec{v}=(v_{x},v_{y},v_{z})$ in the Cartesian
coordinates $\overrightarrow{x}=(x,y,z)$ over homogeneous stationary
background without mean flow $\vec{v}_{0}$ $=0$ , the system (1), (2) looks:

$$\rho_{0}\frac{\partial v_{x}}{\partial t}+\frac{\partial p}{\partial x}%
-\eta\Delta v_{x}-\left(  \varsigma+\frac{\eta}{3}\right)
\frac{\partial }{\partial
x}(\vec{\nabla}v)=-\rho_{0}(\vec{v}\vec{\nabla})v_{x}+\frac
{\rho^{\prime}}{\rho_{0}}\frac{\partial p}{\partial
x}-\eta\rho^{\prime}\Delta v_{x}-\left(
\varsigma+\frac{\eta}{3}\right)  \rho^{\prime}\frac{\partial
}{\partial x}(\vec{\nabla}v)$$

$$\rho_{0}\frac{\partial v_{y}}{\partial t}+\frac{\partial p}{\partial y}%
-\eta\Delta v_{y}-\left(  \varsigma+\frac{\eta}{3}\right)
\frac{\partial }{\partial
y}(\vec{\nabla}v)=-\rho_{0}(\vec{v}\vec{\nabla})v_{y}+\frac
{\rho^{\prime}}{\rho_{0}}\frac{\partial p}{\partial
y}-\eta\rho^{\prime}\Delta v_{y}-\left(
\varsigma+\frac{\eta}{3}\right)  \rho^{\prime}\frac{\partial
}{\partial y}(\vec{\nabla}v)$$

$$\rho_{0}\frac{\partial v_{x}}{\partial t}+\frac{\partial p}{\partial x}%
-\eta\Delta v_{z}-\left(  \varsigma+\frac{\eta}{3}\right)
\frac{\partial }{\partial
z}(\vec{\nabla}v)=-\rho_{0}(\vec{v}\vec{\nabla})v_{z}+\frac
{\rho^{\prime}}{\rho_{0}}\frac{\partial p}{\partial
z}-\eta\rho^{\prime}\Delta v_{z}-\left(
\varsigma+\frac{\eta}{3}\right)  \rho^{\prime}\frac{\partial
}{\partial z}(\vec{\nabla}v)$$

$$\frac{\partial p^{\prime}}{\partial t}+c^{2}\rho_{0}\vec{\nabla}\vec{v}%
-\frac{\chi}{E_{1}}\left(  \frac{\Theta_{1}}{\rho_{0}C_{v}}\Delta p^{\prime
}+\frac{\Theta_{2}p_{0}}{\rho_{0}^{2}C_{v}}\Delta\rho^{\prime}\right)
=\frac{1}{E_{1}}\left(  \varsigma\left(  \vec{\nabla}\vec{v}\right)
^{2}+\frac{\eta}{2}\left(  \frac{\partial v_{i}}{\partial x_{k}}%
+\frac{\partial v_{k}}{\partial
x_{i}}-\frac{2}{3}\delta_{ik}\frac{\partial v_{l}}{\partial
x_{l}}\right)  ^{2}\right)  +\left(  Qp^{\prime}+S\rho
^{\prime}\right)  \vec{\nabla}\vec{v}-\vec{v}\vec{\nabla}p$$

$$\frac{\partial\rho^{\prime}}{\partial t}+\rho_{0}\vec{\nabla}\vec{v}%
=-\rho^{\prime}\vec{\nabla}\vec{v}-\vec{v}\vec{\nabla}\rho^{\prime}
\eqno(4) ,$$

where Q,S are constants depending on the equations of state (2) :

$$Q=\frac{1}{E_{1}}(-1+2\frac{1-E_{2}}{E_{1}}E_{3}+E_{5}),$$ $$S=\frac{1}%
{1-E_{2}}(1+E_{2}+2E_{4}+\frac{1-E_{2}}{E_{1}}E_{5}). \eqno(5)$$

For an ideal gas, $Q=-\gamma=-C_{P}/C_{v},$ $S=0.$\bigskip

In order to simplify calculations, the geometry of beams will be
considered. The equivalent system in the dimensionless variables :
$$\overrightarrow
{v}_{\ast},\overrightarrow{x}_{\ast},\rho_{\ast},p_{\ast},t_{\ast
}:\overrightarrow{v}=\varepsilon
c\overrightarrow{v}_{\ast},p^{\prime
}=\varepsilon c^{2}\rho_{0}p_{\ast},\rho^{\prime}=\varepsilon\rho_{0}%
\rho_{\ast},\overrightarrow{x}=(\lambda
x_{\ast}/\sqrt{\mu},\lambda y_{\ast },\lambda
z_{\ast}/\sqrt{\mu}),t=\lambda t_{\ast}/c,\eqno(6)$$
($c=\sqrt{\frac{p_{0}(1-E_{2})}{\rho_{0}E_{1}}}$is adiabatic sound
velocity, $\lambda$ means characteristic scale of the longitudinal
perturbations, $\mu$ is small parameter expressing the relation
the longitudinal and transverse scales of perturbation,
$\varepsilon$ is small amplitude parameter) looks as follows
(asterisks for dimensionless variables will be later omitted):

$$\frac{\partial}{\partial
t}\psi+L\psi=\varepsilon\varphi+\varepsilon
\varphi_{tv},\eqno(7)$$

where\ $\psi$ is column of field perturbations

$$\psi=\left(
\begin{array}
[c]{ccccc}%
v_{x} & v_{y} & v_{z} & p & \rho
\end{array}
\right)  ^{T},\eqno(8) $$

$L$ is a linear matrix operator :

$$L=\left(
\begin{array}
[c]{ccccc}%
-\delta_{1}^{1}\mu\frac{\partial^{2}}{\partial x^{2}}-\delta_{1}^{2}\Delta &
-\delta_{1}^{1}\sqrt{\mu}\frac{\partial^{2}}{\partial x\partial y} &
-\delta_{1}^{1}\mu\frac{\partial^{2}}{\partial x\partial z} & \sqrt{\mu
}\partial/\partial x & 0\\
-\delta_{1}^{1}\sqrt{\mu}\frac{\partial^{2}}{\partial x\partial y} &
-\delta_{1}^{1}\frac{\partial^{2}}{\partial y^{2}}-\delta_{1}^{2}\Delta &
-\delta_{1}^{1}\sqrt{\mu}\frac{\partial^{2}}{\partial y\partial z} &
\partial/\partial y & 0\\
-\delta_{1}^{1}\mu\frac{\partial^{2}}{\partial x\partial z} & -\delta_{1}%
^{1}\sqrt{\mu}\frac{\partial^{2}}{\partial z\partial y} & -\delta_{1}^{1}%
\mu\frac{\partial^{2}}{\partial z^{2}}-\delta_{1}^{2}\Delta & \sqrt{\mu
}\partial/\partial z & 0\\
\sqrt{\mu}\partial/\partial x & \partial/\partial y & \sqrt{\mu}%
\partial/\partial z & -\delta_{2}^{1}\Delta & -\delta_{2}^{2}\Delta\\
\sqrt{\mu}\partial/\partial x & \partial/\partial y & \sqrt{\mu}%
\partial/\partial z & 0 & 0
\end{array}
\right)\eqno(9) $$

with dimensionless parameters

$\delta_{1}^{1}=\frac{\left(  \zeta+\eta/3\right)  }{\rho_{0}c\lambda}%
,\delta_{1}^{2}=\frac{\eta}{\rho_{0}c\lambda},\delta_{2}^{1}=\frac{\chi
\theta_{1}}{\rho_{0}c\lambda_{0}C_{v}E_{1}},\delta_{2}^{2}=\frac{\chi
\theta_{2}}{\rho_{0}c\lambda_{0}C_{v}(1-E_{2})}.$

Dimensionless operators $\vec{\nabla},\Delta$ $\ $look : $\vec{\nabla}=\left(
\begin{array}
[c]{ccc}%
\mu\partial/\partial x & \partial/\partial y & \mu\partial/\partial z
\end{array}
\right)  ,$ $\Delta=\mu\partial^{2}/\partial x^{2}+\partial^{2}/\partial
y^{2}+\mu\partial^{2}/\partial z^{2},$

$\varphi$ is the quadratic nonlinear column:

\bigskip

\bigskip$$\varphi=\left(
\begin{array}
[c]{c}%
-(\vec{v}\vec{\nabla})v_{x}+\sqrt{\mu}\rho\partial p/\partial x\\
-(\vec{v}\vec{\nabla})v_{y}+\rho\partial p/\partial y\\
-(\vec{v}\vec{\nabla})v_{z}+\sqrt{\mu}\rho\partial p/\partial z\\
\lbrack Qp+S\rho](\vec{\nabla}\vec{v})-(\vec{v}\vec{\nabla})p\\
-\rho(\vec{\nabla}\vec{v})-(\vec{v}\vec{\nabla})\rho
\end{array}
\right) \eqno(10) ,$$

\bigskip and $\varphi_{tv}$ is the quadratic nonlinear column $O(\beta)$
appearing in the viscous flow:

$$\varphi_{tv}=\left(
\begin{array}
[c]{c}%
-\delta_{1}^{2}\rho\Delta v_{x}-\delta_{1}^{1}\rho\frac{\partial}{\partial
x}(\vec{\nabla}v)\\
-\delta_{1}^{2}\rho\Delta v_{y}-\delta_{1}^{1}\rho\frac{\partial}{\partial
y}(\vec{\nabla}v)\\
-\delta_{1}^{2}\rho\Delta v_{z}-\delta_{1}^{1}\rho\frac{\partial}{\partial
z}(\vec{\nabla}v)\\
\frac{1}{E_{1}}\left(  \left(  \delta_{1}^{1}-\delta_{1}^{2}/3\right)  \left(
\vec{\nabla}\vec{v}\right)  ^{2}+\frac{\delta_{1}^{2}}{2}\left(
\frac{\partial v_{i}}{\partial x_{k}}+\frac{\partial v_{k}}{\partial x_{i}%
}-\frac{2}{3}\delta_{ik}\frac{\partial v_{l}}{\partial x_{l}}\right)
^{2}\right) \\
0
\end{array}
\right) \eqno(11) $$
For a linear flow defined by the linearized\ version of system (7)

$$\frac{\partial}{\partial t}\psi+L\psi=0,\eqno(12)$$ a solution may
be found as a sum of planar waves: $\ v_{x}=\widetilde
{v_{x}}(\overrightarrow{k})\exp(i\omega
t-i\overrightarrow{k}\overrightarrow {x}),....$where

$$\overrightarrow{k}=(k_{x},k_{y},k_{z})$$ is the wave vector. It is
convenient to go to the Fourier transforms marked by tilde for the
flows over the homogeneous background. In the Fourier
space,$-ik_{x}$ means $\partial/\partial x,$ $i\omega$ means
$\partial/\partial t,$and (6) yields in
the five roots of dispersion relation $(\beta=\delta_{1}^{1}+\delta_{1}%
^{2}+\delta_{2}^{1}+\delta_{2}^{2})$:

$\omega_{1}=\Omega+i\beta\Omega^{2}/2,$ \ \
$\omega_{2}=-\Omega+i\beta \Omega^{2}/2,$ \ \
$\omega_{3}=-i\delta_{2}^{2}\Omega^{2},$\ $\ \omega
_{4}=i\delta_{1}^{2}\Omega^{2},$
$\omega_{5}=i\delta_{1}^{2}\Omega^{2},$where
$$\Omega=k_{y}+\frac{\mu(k_{x}^{2}+k_{z}^{2})}{2k_{y}}.\eqno(13)$$
Also, an operator $\left(  -ik_{y}\right)  ^{-1}$ represents in
$\overrightarrow{k}$-space operator $\int dy$ . Constant of
integration should be chosen accordingly to the concrete physical
problem. The first two roots relate to progressive (acoustic)
modes of different directions of propagation, the third one
relates to the entropy mode, and the fourth and fifth ones - to
the vortical ones. For the real substances, $\beta>0,$and, $\delta_{2}^{2}%
<0,$that corresponds physically correct sign of imagine parts of all
frequencies. The modes of linear flow are determined by relations of
amplitudes of plane waves $\widetilde{v_{x}}(k_{x},k_{y},k_{z})$ ,$\ldots$
Modes as eigenvectors of a linear problem in the Fourier space look:

\bigskip$\widetilde{\psi}_{1}=\left(
\begin{array}
[c]{c}%
\widetilde{v_{x}}_{1}(k_{x},k_{y},k_{z})\\
\widetilde{v_{y}}_{1}(k_{x},k_{y},k_{z})\\
\widetilde{v_{z}}_{1}(k_{x},k_{y},k_{z})\\
\widetilde{p}_{1}(k_{x},k_{y},k_{z})\\
\widetilde{\rho}_{1}(k_{x},k_{y},k_{z})
\end{array}
\right)  =\left(
\begin{array}
[c]{c}%
\sqrt{\mu}k_{x}/k_{y}\\
1-\mu(k_{x}^{2}+k_{z}^{2})/(2k_{y}^{2})+i\beta k_{y}/2\\
\sqrt{\mu}k_{z}/k_{y}\\
1+i(\delta_{2}^{1}+\delta_{2}^{2})k_{y}\\
1
\end{array}
\right)  \widetilde{\rho}_{1}=\widetilde{M}_{1}\widetilde{\rho}_{1},$

$$\widetilde{\psi}_{2}=\left(
\begin{array}
[c]{c}%
-\sqrt{\mu}k_{x}/k_{y}\\
-1+\mu(k_{x}^{2}+k_{z}^{2})/(2k_{y}^{2})+i\beta k_{y}/2\\
-\sqrt{\mu}k_{z}/k_{y}\\
1-i(\delta_{2}^{1}+\delta_{2}^{2})k_{y}\\
1
\end{array}
\right)
\widetilde{\rho}_{2}=\widetilde{M}_{2}\widetilde{\rho}_{2},\eqno(14)$$
$\widetilde{\psi}_{3}=\left(
\begin{array}
[c]{c}%
0\\
-i\delta_{2}^{2}k_{y}\\
0\\
0\\
1
\end{array}
\right)  \widetilde{\rho}_{3}=\widetilde{M}_{3}\widetilde{\rho}_{3},$
$\ \widetilde{\psi}_{4}=\left(
\begin{array}
[c]{c}%
ik_{y}\\
-i\sqrt{\mu}k_{x}\\
0\\
0\\
0
\end{array}
\right)  \widetilde{\varphi}_{4}=\widetilde{M}_{4}\widetilde{\varphi}%
_{4},\ \widetilde{\psi}_{5}=\left(
\begin{array}
[c]{c}%
0\\
-i\sqrt{\mu}k_{z}\\
ik_{y}\\
0\\
0
\end{array}
\right)  \widetilde{\varphi}_{5}=\widetilde{M}_{5}\widetilde{\varphi}_{5}.$

All calculations of modes and projectors have accuracy up to the terms of
order $\mu$, $\beta$. As basic variable for the first three modes the
perturbation of density is chosen, and for the last two- the stream function,
since for the both vorticity modes density keeps unperturbed. So, there are
five independent values indeed, one for every mode, that define other wave
perturbations of the mode uniquely. Any field of the linear flow as a solution
of a linearized equation (12) may be presented as a sum of independent modes.

The next step is to get projectors that decompose a concrete mode from the
overall field $\widetilde{\psi}.$ Let us define a matrix M in the following way:

$$\widetilde{M}\left(
\begin{array}
[c]{c}%
\widetilde{\rho}_{1}\\
\widetilde{\rho}_{2}\\
\widetilde{\rho}_{3}\\
\widetilde{\varphi}_{4}\\
\widetilde{\varphi}_{4}%
\end{array}
\right)
=\widetilde{\psi}=\underset{i=1}{\overset{5}{\sum}}\widetilde{\psi
}_{i}=\widetilde{M}_{1}\widetilde{\rho}_{1}+\widetilde{M}_{2}\widetilde{\rho
}_{2}+\widetilde{M}_{3}\widetilde{\rho}_{3}+\widetilde{M}_{4}\widetilde
{\varphi}_{4}+\widetilde{M}_{5}\widetilde{\varphi}_{5}.\eqno(15)$$

so that matrix $\widetilde{M}$ consists of five columns,

$$\widetilde{M}=(%
\begin{array}
[c]{ccccc}%
\widetilde{M}_{1} & \widetilde{M}_{2} & \widetilde{M}_{3} & \widetilde{M}_{4}
& \widetilde{M}_{5}%
\end{array}
).\eqno(16)$$ and the inverse matrix $\widetilde{M}^{-1}$ consists
of five lines

$\widetilde{M}^{-1}=\left(
\begin{array}
[c]{ccccc}%
\widetilde{M}_{1}^{-1} & \widetilde{M}_{2}^{-1} & \widetilde{M}_{3}^{-1} &
\widetilde{M}_{4}^{-1} & \widetilde{M}_{5}^{-1}%
\end{array}
\right)  ^{T},$

where lines $\widetilde{M}_{1}^{-1},..\widetilde{M}_{5}^{-1}$ are as follows:

$\widetilde{M}_{1}^{-1}=(%
\begin{array}
[c]{ccccc}%
\sqrt{\mu}\frac{k_{x}}{k_{y}} & \frac{1}{2}\left(  1-i(\delta_{2}^{1}%
+\delta_{2}^{2})k_{y}-\mu(k_{x}^{2}+k_{z}^{2})/(2k_{y})\right)  & \sqrt{\mu
}\frac{k_{z}}{k_{y}} & \frac{1}{2}\left(  1-i\beta k_{y}/2-i\delta_{2}%
^{2}k_{y}\right)  & i\delta_{2}^{2}k_{y}/2
\end{array}
)$,

$\widetilde{M}_{2}^{-1}=(%
\begin{array}
[c]{ccccc}%
-\sqrt{\mu}\frac{k_{x}}{k_{y}} & \frac{1}{2}\left(  -1-i(\delta_{2}^{1}%
+\delta_{2}^{2})k_{y}+\mu(k_{x}^{2}+k_{z}^{2})/(2k_{y})\right)  & -\sqrt{\mu
}\frac{k_{z}}{k_{y}} & \frac{1}{2}\left(  1+i\beta k_{y}/2+i\delta_{2}%
^{2}k_{y}\right)  & -i\delta_{2}^{2}k_{y}/2
\end{array}
),$

$\widetilde{M}_{3}^{-1}=(%
\begin{array}
[c]{ccccc}%
0 & \frac{1}{2}i(\delta_{2}^{1}+\delta_{2}^{2})k_{y} & 0 & -1 & 1
\end{array}
),$

$\widetilde{M}_{4}^{-1}=(%
\begin{array}
[c]{ccccc}%
i\frac{1-\mu k_{x}^{2}/k_{y}^{2}}{k_{y}} & -i\sqrt{\mu}\frac{k_{x}}{k_{y}^{2}}
& -i\mu\frac{k_{x}k_{z}}{k_{y}^{3}} & 0 & 0
\end{array}
),$

$\widetilde{M}_{5}^{-1}=(%
\begin{array}
[c]{ccccc}%
-i\mu\frac{k_{x}k_{z}}{k_{y}^{3}} & -i\sqrt{\mu}\frac{k_{z}}{k_{y}^{2}} &
i\frac{1-\mu k_{z}^{2}/k_{y}^{2}}{k_{y}} & 0 & 0
\end{array}
).$$$\eqno(17)$$

Accordingly to definition of matrix $\widetilde{M}^{-1}$ projectors may be
determined, so as

$\widetilde{P}_{1}\widetilde{\psi}=\widetilde{\psi}_{1},...,\widetilde{P}%
_{5}\widetilde{\psi}=\widetilde{\psi}_{5}.$ \ \ \ \ \ \ \ \ \ \ \ \ \ \ \ \ \ \ \ \ \ \ \ \ \ \ \ \ \ \ \ \ \ \ \ \ \ \ \ \ \ \ \ \ \ \ \ \ \ \ \ \ \ \ \ \ \ \ \ \ \ \ \ \ \ \ \ \ \ \ \ \ \ \ \ \ \ \ \ \ \ \ \ \ \ \ \ \ \ \ \ \ \ \ \ \ \ \ \ \ \ \ \ \ \ \ \ \ \ \ \ \ \ \ \ \ \ \ \ \ \ \ \ \ \ \ \ \ \ \ \ \ \ \ \ \ \ \ \ \ \ \ \ \ \ \ \ \ \ \ \ (18)

Relations (14), (15) written for the first mode as an example $\widetilde
{M}_{1}^{-1}\widetilde{\psi}=\rho_{1},\widetilde{\psi}_{1}=\widetilde{M}%
_{1}\rho_{1}$ lead to $\widetilde{P}_{1}=\widetilde{M}_{1}\cdot\widetilde
{M}_{1}^{-1}$ accordingly to (18) and so on for the all other modes:

$$\widetilde{P}_{i}=\widetilde{M}_{i}\cdot\widetilde{M}_{i}^{-1},i=1,..5.\eqno(19)$$

At any moment of evolution a concrete mode is distinguished from the overall
field by the correspondent projector. Projectors calculated with accuracy of
order $\mu,\beta$ look:

$\widetilde{P}_{1,2}=\left(
\begin{array}
[c]{ccccc}%
\mu\frac{k_{x}^{2}}{2k_{y}^{2}} & \sqrt{\mu}\frac{k_{x}}{2k_{y}} & \mu
\frac{k_{x}k_{z}}{2k_{y}^{2}} & \pm\sqrt{\mu}\frac{k_{x}}{2k_{y}} & 0\\
\sqrt{\mu}\frac{k_{x}}{2k_{y}} & \frac{1}{2}\left(  1\pm\frac{i\beta}{2}%
k_{y}\mp i(\delta_{2}^{1}+\delta_{2}^{2})k_{y}-\mu\frac{k_{x}^{2}+k_{z}^{2}%
}{2k_{y}^{2}}\right)  & \sqrt{\mu}\frac{k_{z}}{2k_{y}} & \frac{1}{2}\left(
\pm1-i\delta_{2}^{2}k_{y}-\mu\frac{k_{x}^{2}+k_{z}^{2}}{2k_{y}^{2}}\right)  &
\frac{i\delta_{2}^{2}k_{y}}{2}\\
\mu\frac{k_{x}k_{z}}{2k_{y}^{2}} & \sqrt{\mu}\frac{k_{z}}{2k_{y}} & \mu
\frac{k_{z}^{2}}{2k_{y}^{2}} & \pm\sqrt{\mu}\frac{k_{z}}{2k_{y}} & 0\\
\pm\sqrt{\mu}\frac{k_{x}}{2k_{y}} & \pm\frac{1}{2}\left(  1-\mu\frac{k_{x}%
^{2}+k_{z}^{2}}{2k_{y}^{2}}\right)  & \pm\sqrt{\mu}\frac{k_{z}}{2k_{y}} &
\frac{1}{2}(1\mp\frac{i\beta}{2}k_{y}\pm i\delta_{2}^{1}k_{y}) & \pm
\frac{i\delta_{2}^{2}k_{y}}{2}\\
\pm\sqrt{\mu}\frac{k_{x}}{2k_{y}} & \frac{1}{2}\left(  \pm1-i(\delta_{2}%
^{1}+\delta_{2}^{2})k_{y}\mp\mu\frac{k_{x}^{2}+k_{z}^{2}}{2k_{y}^{2}}\right)
& \pm\sqrt{\mu}\frac{k_{z}}{2k_{y}} & \frac{1}{2}(1\mp\frac{i\beta}{2}k_{y}\mp
i\delta_{2}^{2}k_{y}) & \pm\frac{i\delta_{2}^{2}k_{y}}{2}%
\end{array}
\right)  ,$

$$\widetilde{P}_{3}=\left(
\begin{array}
[c]{ccccc}%
0 & 0 & 0 & 0 & 0\\
0 & 0 & 0 & i\delta_{2}^{2}k_{y} & -i\delta_{2}^{2}k_{y}\\
0 & 0 & 0 & 0 & 0\\
0 & 0 & 0 & 0 & 0\\
0 & i(\delta_{2}^{1}+\delta_{2}^{2})k_{y} & 0 & -1 & 1
\end{array}
\right)  ,\eqno(20)$$

$\widetilde{P}_{4}=\left(
\begin{array}
[c]{ccccc}%
1-\mu\frac{k_{x}^{2}}{k_{y}^{2}} & -\sqrt{\mu}\frac{k_{x}}{k_{y}} & -\mu
\frac{k_{x}k_{z}}{k_{y}^{2}} & 0 & 0\\
-\sqrt{\mu}\frac{k_{x}}{k_{y}} & \mu\frac{k_{x}^{2}}{k_{y}^{2}} & 0 & 0 & 0\\
0 & 0 & 0 & 0 & 0\\
0 & 0 & 0 & 0 & 0\\
0 & 0 & 0 & 0 & 0
\end{array}
\right)  ,\widetilde{P}_{5}=\left(
\begin{array}
[c]{ccccc}%
0 & 0 & 0 & 0 & 0\\
0 & \mu\frac{k_{z}^{2}}{k_{y}^{2}} & -\sqrt{\mu}\frac{k_{z}}{k_{y}} & 0 & 0\\
-\mu\frac{k_{x}k_{z}}{k_{y}^{2}} & \sqrt{\mu}\frac{k_{z}}{k_{y}} & 1-\mu
\frac{k_{z}^{2}}{k_{y}^{2}} & 0 & 0\\
0 & 0 & 0 & 0 & 0\\
0 & 0 & 0 & 0 & 0
\end{array}
\right)  ,.$

Matrix projectors satisfy common properties of orthogonal projectors:

$\overset{5}{\underset{i=1}{\sum}}\widetilde{P}_{i}=\widetilde{I},$
$\widetilde{P}_{i}\cdot\widetilde{P}_{n}=\widetilde{0}$ if $i\neq
n,\widetilde{P}_{i}\cdot\widetilde{P}_{i}=$ $\widetilde{P}_{i}$ $if$ $\ i=n,$ \ \ \ \ \ \ \ \ \ \ \ \ \ \ \ \ \ \ \ \ \ \ \ \ \ \ \ \ \ \ \ \ \ \ \ \ \ \ \ \ \ \ \ \ \ \ \ \ \ \ \ \ \ \ \ \ \ \ \ \ \ \ \ \ \ \ \ \ \ \ \ \ \ \ \ \ \ \ \ \ \ \ \ \ \ \ \ \ \ \ \ \ \ \ \ \ \ \ \ \ \ \ \ \ \ \ \ \ \ (21)

where $\widetilde{I},\widetilde{0}$ are unit and zero matrices. The inverse
transformation of formulae (20) to the $(\overrightarrow{x},t)$ space may be
easily undertaken.

\section{ Linear flow and evolution equations.}

A linear flow may be decomposed to modes uniquely accordingly to (18). Linear
evolution equations for every mode may be originated from the linear version
of (7) by acting of the corresponding projector: $P\left(  \frac{\partial
}{\partial t}\psi+L\psi\right)  =0.$ Note that all projectors do commute with
both $\partial/\partial t$ $\cdot I$ and $L$ ($I$ is unit matrix). When one
acts by projector at the linear system (12), five equations for all the
components of every mode appear. For example, an evolution equation for the
dimensionless perturbations of density for both acoustic modes are as follows:

$\frac{\partial\rho_{1,2}}{\partial
t}\pm\frac{\partial\rho_{1,2}}{\partial
y}\pm\frac{\mu}{2}\int\Delta_{\perp}\rho_{1,2}dy-\frac{\beta}{2}\frac
{\partial^{2}\rho_{1,2}}{\partial y^{2}}=0,$ \ \ \ \ \ \ \ \ \ \ \
\ \ \ \ \ \ \ \ \ \ \ \ \ \ \ \ \ \ \ \ \ \ \ \ \ \ \ \ \ \ \ \ \
\ \ \ \ \ \ \ \ \ \ \ \ \ \ \ \ \ \ \ \ \ \ \ \ \ \ \ \ \ \ \ \ \
\ \ \ \ \ \ \ \ \ \ \ \ \ \ \ \ \ \ \ \ \ \ \ \ \ \ \ \ \ \ \ \ \
\ \ \ \ \ \ \ \ \ \ \ \ \ (22)

where $\Delta_{\perp}=\frac{\partial^{2}}{\partial x^{2}}+\frac{\partial^{2}%
}{\partial z^{2}}.$

An evolution equation for the axial velocity of the entropy mode looks:

$\frac{\partial v_{y3}}{\partial t}-\delta_{2}^{2}\frac{\partial^{2}v_{y3}%
}{\partial y^{2}}=0,$ \ \ \ \ \ \ \ \ \ \ \ \ \ \ \ \ \ \ \ \ \ \ \ \ \ \ \ \ \ \ \ \ \ \ \ \ \ \ \ \ \ \ \ \ \ \ \ \ \ \ \ \ \ \ \ \ \ \ \ \ \ \ \ \ \ \ \ \ \ \ \ \ \ \ \ \ \ \ \ \ \ \ \ \ \ \ \ \ \ \ \ \ \ \ \ \ \ \ \ \ \ \ \ \ \ \ \ \ \ \ \ \ \ \ \ \ \ \ \ \ \ \ \ \ \ \ \ \ \ \ \ \ \ \ \ \ \ \ \ \ \ \ \ \ \ \ \ \ \ \ \ \ \ \ \ \ \ \ \ \ \ \ \ (23)

and for the axial velocity of two vortical modes is:

$\frac{\partial v_{y4,5}}{\partial t}-\delta_{1}^{2}\frac{\partial^{2}%
v_{y4,5}}{\partial y^{2}}=0.$ \ \ \ \ \ \ \ \ \ \ \ \ \ \ \ \ \ \ \ \ \ \ \ \ \ \ \ \ \ \ \ \ \ \ \ \ \ \ \ \ \ \ \ \ \ \ \ \ \ \ \ \ \ \ \ \ \ \ \ \ \ \ \ \ \ \ \ \ \ \ \ \ \ \ \ \ \ \ \ \ \ \ \ \ \ \ \ \ \ \ \ \ \ \ \ \ \ \ \ \ \ \ \ \ \ \ \ \ \ \ \ \ \ \ \ \ \ \ \ \ \ \ \ \ \ \ \ \ \ \ \ \ \ \ \ \ \ \ \ \ \ \ \ \ \ \ \ \ \ \ \ \ \ \ \ \ \ \ \ \ \ (24)

These linear evolution equations are well-known .Equations (21)-(23) are
explicit up to terms of order $O(\beta\sqrt{\mu},\mu^{2})$ due to accuracy of
evaluated projectors. In the paper [9], equations (22)-(24) are presented as
three equations for the three basic modes, since there was no subdivision of
acoustic and vortical modes into two independent branches everyone. Also, the
evolution equations from [9] involves sources of mass, momentum and heat that
easily may be added in to the right-hand side of the initial system (7) as
well as to the equations (22)-(24).

\bigskip

\section{Nonlinear flow and coupled evolution equations.}

At this point, we declare that relations inside every mode become fixed in
weakly nonlinear flow. Acting by projectors at the original system (7) with
non-zero nonlinear part leads to coupled equations for interacting modes:
$P\left(  \frac{\partial}{\partial t}\psi+L\psi\right)  =\varepsilon P\left(
\varphi_{1}+\varphi_{1tv}\right)  $. Modes become separated in the linear
side, and one should account that the overall field is a sum of all modes
inputs in the nonlinear side. The proper returning to the $\overrightarrow{x}
$ space should be proceeded as well. To illustrate the possibilities of the
projecting method, some illustrations concerning the famous evolution
equations deriving are presented below.

\subsection{One-dimensional nonlinear flow.}

Let $k_{x}=0,$ $k_{z}=0,$ that corresponds to one-dimensional flow along
y-axis. For the planar geometry, only three modes exist: two acoustic and the
entropy one. Acting by $P_{1}$ at the system (5), one get an evolution
equation for the rightward acoustic mode consisting of three equations for ,
$p_{1},\rho_{1},$ and $v_{1}$. An evolution equation for the non-dimensional
density perturbations of the rightward acoustic mode is as follows:

$$\frac{\partial\rho_{1}}{\partial t}+\frac{\partial\rho_{1}}{\partial y}%
-\frac{\beta}{2}\frac{\partial^{2}v_{1}}{\partial y^{2}}=\frac{\varepsilon}%
{2}\left(  -v\frac{\partial}{\partial
y}v+\rho\frac{\partial}{\partial y}p-v\frac{\partial}{\partial
y}p+\left[  Qp+S\rho\right]  \frac{\partial }{\partial y}v\right)
\eqno(25) ,$$ Here, $p,\rho,$ and $v$ in the right-hand part
represent a sum of all modes:

$v=v_{1}+v_{2}+v_{3},$ $p=p_{1}+p_{2}+p_{3},$
$$\rho=\rho_{1}+\rho_{2}+\rho _{3}.\eqno(26)$$ All cross
nonlinear-viscous terms are neglected, as well as higher order
nonlinear ones. If the nonlinear terms of the first mode are kept
only, we get the known Burgers evolution equation, going to the
famous Earnshow one in the limit $\beta\rightarrow0$ :

$$\frac{\partial\rho_{1}}{\partial t}+\frac{\partial\rho_{1}}{\partial
y}+\varepsilon\frac{-Q-S+1}{2}\rho_{1}\frac{\partial}{\partial
y}\rho _{1}-\frac{\beta}{2}\frac{\partial^{2}\rho_{1}}{\partial
y^{2}}=0.\eqno(27)$$

Physically, this case relates to a dominant rightward acoustic mode. For an
ideal gas, again, $Q=-\gamma=-C_{P}/C_{v},$

$S=0.$\bigskip

\subsection{Dynamics of beams.}

In the same way, the famous Khokhlov-Zabolotskaya equation [14,15] (KZ, in the
limit $\beta\rightarrow0$) and Khokhlov-Zabolotskaya-Kuznetsov equation (KZK)
for rightward (leftward) propagating beams follows when acting by $P_{1,}%
P_{2}$ at the system (7):

$$\frac{\partial\rho_{1,2}}{\partial t}\pm\frac{\partial\rho_{1,2}}{\partial
y}\pm\frac{\mu}{2}\int\Delta_{\perp}\rho_{1,2}dy-\frac{\beta}{2}\frac
{\partial^{2}\rho_{1,2}}{\partial y^{2}}\pm\varepsilon\frac{-Q-S+1}{2}%
\rho_{1,2}\frac{\partial}{\partial y}\rho_{1,2}=0,\eqno(28)$$

This evolution equation may be understood as a limit for the progressive mode
self-action, meaning that only this mode nonlinear terms are kept. It seems be
suitable definition though the 'self-action' is usually reserved for the
phenomenon of acoustic heating when acoustic mode causes the entropy mode
during its propagation thus changing the surrounding [8].

\bigskip

\section{ACOUSTIC STREAMING CAUSED BY IMPULSES.}

There are some reasons to find another approach to acoustic
streaming. First, the procedure of temporal averaging is a basis
of the modern theory of acoustic streaming and all corresponding
results relate to averaged fields. All non-periodic acoustic
sources thus are left of account. Second, the problem of modes
interaction looks more extended. Actually there exist three types
of independent modes. The initial amplitudes of these modes are
determined by initial conditions of a concrete problems and may be
evaluated by projecting of the overall initial field into every
modal field. The acoustic streaming imposes an ability of
predominate acoustic mode. The acoustic field causes slow varying
with time flow which may grow with time since nonlinear effects
can store, that is namely the acoustic streaming. So, the acoustic
streaming relates to one type (though important) of possible
nonlinear interactions in flow, namely to the beginning of
acoustic mode evolution. Later, an amplitude of the vortical mode
(and the entropy one also) may become so large that one can not
treat these modes as secondary and other approach for nonlinear
interaction of modes should be developed. Let us mark, that even
traditional subdivision of overall flow into 'slow' and 'quick'
components eliminates the entropy mode unclearly. It is also
'slow' but essentially possesses non-zero density perturbation
which slowly varies with time ( $\partial\rho_{3}/\partial t=0$ in
linear approach), so the procedure of temporal averaging of
continuity equation fails with storing of nonlinear effects. Also,
there are many more complicated flows over inhomogeneous media
and/or with background flows that may be algorithmically solved by
the pointed method, see conclusion.

We will show, at first, that projecting results in the well-known equations
for an acoustic streaming in terms of $\overrightarrow{V},$ and, the second
and more important, that there appear the new equations for acoustic streaming
caused by impulse (non-periodic) acoustic signals.

To consider interactions of vortical and acoustic modes, one acts by projector
$P_{4}$ $+$ $P_{5}$ at the system (7) to get an evolution equation for
vorticity $\varphi_{4}$ $+$ $\varphi_{5}$ . The right-hand nonlinear vector
involves, again, all modes inputs. Then, to calculate the nonlinear
interactions of only vortical and rightward propagating acoustic mode, the
right-hand vector should be thought as a sum of vortical and the first
acoustic mode:

$$\psi=\left(
\begin{array}
[c]{c}%
\sqrt{\mu}\frac{\partial}{\partial x}\int dy\\
1-\frac{\mu}{2}\Delta_{\perp}\int dy\int dy-\frac{\beta}{2}\frac{\partial
}{\partial y}\\
\sqrt{\mu}\frac{\partial}{\partial z}\int dy\\
1-(\delta_{2}^{1}+\delta_{2}^{2})\frac{\partial}{\partial y}\\
1
\end{array}
\right)  \rho_{1}+\left(
\begin{array}
[c]{c}%
v_{x4}\\
v_{y4}\\
0\\
0\\
0
\end{array}
\right)  +\left(
\begin{array}
[c]{c}%
0\\
v_{y5}\\
v_{z5}\\
0\\
0
\end{array}
\right)  .\eqno(29)$$

An operator $\Delta_{\perp}\int dy\int dy$ represents in $\overrightarrow{x}%
$-space operator $\frac{k_{x}^{2}+k_{z}^{2}}{k_{y}^{2}}.$Acting by
$P_{4}+P_{5}$ at the system (5) leads to an evolution equation (27) for
vorticity $\varphi(\overrightarrow{x},t),$ which we rewrite going to velocity
components $V_{x}=v_{x4}=-\partial\varphi_{4}/\partial y,$ $V_{y}%
=v_{y4}+v_{y5}=\sqrt{\mu}\partial\varphi_{4}/\partial x+\sqrt{\mu}%
\partial\varphi_{5}/\partial z,$ $V_{z}=v_{z5}=-\partial\varphi_{5}/\partial
y:$

$$\frac{\partial V_{x}}{\partial t}-\delta_{1}^{2}\frac{\partial
^{2}V_{x}}{\partial y^{2}}=-\varepsilon(\overrightarrow{V}%
\overrightarrow{\nabla})V_{x}+\varepsilon F_{1x}\eqno(30)$$,

where in right-hand side of (30) only two term are left: the first relating to
the vortical mode, and the second $F_{1}$ consisting of quadratic inputs of
the first progressive acoustic mode:\ \

\bigskip$$F_{1x}=\left(
\begin{array}
[c]{ccccc}%
1-\mu\frac{\partial^{2}}{\partial x^{2}}\int dy\int dy & -\sqrt{\mu}%
\frac{\partial}{\partial x}\int dy & -\mu\frac{\partial^{2}}{\partial
x\partial z}\int dy\int dy & 0 & 0
\end{array}
\right)  (\varphi_{1}+\varphi_{1tv}),\eqno(31)$$

where $\varphi_{1},\varphi_{1tv}$ are supposed to consist of specific
perturbations for the rightward dominant acoustic mode, see (10). For example,
$\varphi_{1}$ looks

\bigskip$\varphi_{1}=\left(
\begin{array}
[c]{c}%
-(\vec{v}_{1}\vec{\nabla})v_{1x}+\sqrt{\mu}\rho_{1}\partial p_{1}/\partial x\\
-(\vec{v}_{1}\vec{\nabla})v_{1y}+\rho_{1}\partial p_{1}/\partial y\\
-(\vec{v}_{1}\vec{\nabla})v_{1z}+\sqrt{\mu}\rho_{1}\partial p_{1}/\partial z\\
\lbrack Qp_{1}+S\rho_{1}](\vec{\nabla}\vec{v}_{1})-(\vec{v}_{1}\vec{\nabla
})p_{1}\\
-\rho_{1}(\vec{\nabla}\vec{v}_{1})-(\vec{v}_{1}\vec{\nabla})\rho_{1}%
\end{array}
\right)  .$ \ There are all other possible cross nonlinear terms in the
right-hand side of (30) which are out of interest in the present investigation
since the rightward acoustic mode supposed to be dominant at least at the
initial stage of the evolution.\ \ \ Writing on the terms of order not higher
then $\beta\mu$ , one goes to the expected result:

$$\left(
\begin{array}
[c]{ccccc}%
1-\mu\frac{\partial^{2}}{\partial x^{2}}\int dy\int dy & -\sqrt{\mu}%
\frac{\partial}{\partial x}\int dy & -\mu\frac{\partial^{2}}{\partial
x\partial z}\int dy\int dy & 0 & 0
\end{array}
\right)  \varphi_{1}=0,\eqno(32)$$ that means that\ acoustical
streaming may exist only in the thermoviscous flow. Finally,\ \
$F_{1x}$ is as follows

$F_{1x}=\left(
\begin{array}
[c]{ccccc}%
1-\mu\frac{\partial^{2}}{\partial x^{2}}\int dy\int dy & -\sqrt{\mu}%
\frac{\partial}{\partial x}\int dy & -\mu\frac{\partial^{2}}{\partial
x\partial z}\int dy\int dy & 0 & 0
\end{array}
\right)  \varphi_{1tv}=$

$\sqrt{\mu}\left[  \frac{\beta}{2}\left(  \frac{\partial\rho_{1}}{\partial
x}\frac{\partial\rho_{1}}{\partial y}-\frac{\partial}{\partial x}\int
dy(\rho_{1}\frac{\partial^{2}\rho_{1}}{\partial y^{2}})-\frac{\partial
}{\partial x}\int dy(\frac{\partial\rho_{1}}{\partial y})^{2}\right)
+(\delta_{2}^{1}+\delta_{2}^{2})\left(  \frac{\partial}{\partial x}\int
dy(\rho_{1}\frac{\partial^{2}\rho_{1}}{\partial y^{2}})-\rho_{1}\frac
{\partial^{2}\rho_{1}}{\partial x\partial y}\right)  \right]  +$

$$\sqrt{\mu}\left[  \beta\left(  \frac{\partial}{\partial x}\int dy\left(
\rho_{1}\frac{\partial^{2}\rho_{1}}{\partial y^{2}}\right)  -\rho_{1}%
\frac{\partial^{2}\rho_{1}}{\partial x\partial y}\right)  +(\delta_{1}%
^{1}+\delta_{1}^{2})\left(  \frac{\partial}{\partial x}\int dy(\rho_{1}%
\frac{\partial^{2}\rho_{1}}{\partial y^{2}})-\rho_{1}\frac{\partial^{2}%
\rho_{1}}{\partial x\partial y}\right)  \right]  . \eqno(33)$$

\ To show that both effects play role: (1) presence of the thermoviscous terms
in $\psi_{1},$ and (2) presence of thermoviscous terms in $\varphi_{1tv},$
corresponding terms are included in square brackets, first and second
ones\ though they partially compensate each other. Finally, (33) goes to the following:

$$F_{1x}=\sqrt{\mu}\beta\left(  \frac{\partial\rho_{1}}{\partial x}%
\frac{\partial\rho_{1}}{\partial y}-0.5\rho_{1}\frac{\partial^{2}\rho_{1}%
}{\partial x\partial y}-\frac{\partial}{\partial x}\int
dy(\frac{\partial \rho_{1}}{\partial y})^{2}\right)  ,\eqno(34)$$

and the constant of integration of the last term should be chosen
accordingly to the real initial perturbations and /or boundary
regime. It is remarkable that only three first elements of
$\varphi_{1tv}$ participate in the final radiation force, namely
that three ones are responsible for the losses of momentum in the
flow. The forth element expresses the thermoviscous loss of
energy. This term (and also the fifth one) plays role in the
forming of the heating, see the last line of $\widetilde{P}_{3}$
given by (20). Relations of the specific perturbations of
eigenvector $\psi_{1}$ (14) are used to express $F_{1x}$ through
$\rho_{1}$.

The transverse component of the force caused by sound $F_{1x}$ given by
formula (34) does not possesses terms depending on the other transverse
coordinate z because these terms are of order $\beta\mu^{3/2}$ . In all
calculations, the properties of specific modes and related projectors applied
only. The temporal averaging was not proceeded. It has been proved that the
acoustic force should be rotational [16]. In the referred papers, the
rotational part of the averaged is just selected from the general expression.
Projectors $P_{4,}P_{5}$ due to their properties yield in the vortical forces automatically.

\bigskip\

\subsection{Limit of the quasi-periodic acoustic source}

To compare formula (30) with this caused by periodic at the tranducer source,
let us take the transverse acoustic force (radiation force) given by Gusev,
Rudenko [17]

$$\Phi_{1x}= \sqrt{\mu}\rho_{0}^{2}\frac{\partial}{\partial
x}(\theta^{2}/2)\exp(-\beta y)\eqno(35)$$

rewritten in our variables; note also that longitudinal coordinate there is
$x$ instead of $y$. The radiation force (35) relates to the streaming evolving
in the sound field of a highly attenuated beam with a quasi-periodic source as follows:

$$\rho_{1}(x,y,t)=\rho_{0}\theta(x)\exp(-\frac{\beta}{2}y)\sin(t-y)\eqno(36)$$
see [12] for more details.\ Calculating based on formula (34)
yields in the following:

$$F_{1x}=\sqrt{\mu}\rho_{0}^{2}\frac{\partial}{\partial
x}(\theta^{2}/2)\exp(-\beta y)
\left[1+\frac{\beta}{4}\sin(2(t-y))+O(\beta ^{2})\right] \eqno(37)
$$

Averaging over\ integral number of period of the sound wave in the leading
order gives $\left\langle F_{1x}\right\rangle =\Phi_{1x}.$ \ In contrast to
the known results, (34) fits to calculate force caused by any acoustic source
including non-periodic ones. Moreover, evolution of streaming given by
temporal averaging is not suitable in principle for detail tracing with small
temporal step because of the very initial procedure of averaging over time
whose interval is much longer than a period of a sound wave.

\subsection{Radiation force cased by mono-polar acoustic source}

 As an appropriate acoustic source satisfying the limitations
of [17], let us take a mono-polar\ two-dimensional wave as follows [18]:

$$\rho_{1}(x,y,t)=-\sqrt{\frac{2\beta}{\pi}}\exp(-x^{2})\frac{\exp\left(
-\tau^{2}/2\xi\right)  }{\epsilon\sqrt{\xi/\beta}\left(
C-Erf(\tau/\sqrt {2\xi}\right)  },\eqno(38)  $$ where $\xi=\beta
y$, \ $\tau=t-y,$ $\epsilon=\frac{-Q-S+1}{2}$ is a parameter of
nonlinearity, see (5). For ideal gases,
$\epsilon=\frac{\gamma+1}{2}.$ Constant C is responsible for the
shape of the curve: large C provides a curve close to the Gauss
one. The function $F_{1x}$ achieves negative minimum at the
transversal distance $x_{m}=\sqrt{2}/2$ and is an odd function as
follows from (34), (38). Results of the calculating of
$\frac{F_{1x}}{\sqrt{\mu}\beta}$ $\ $\ in accordance to formula
(34) are shown at the figures 1,2 at the transversal point
$x=x_{m}$. Figure 1 shows the dependence of
$\frac{F_{1x}}{\sqrt{\mu}\beta}$ on $y$ at $t=1$, and figure 2
shows the same at $t=3.$ Constants of mono-polar source (35) used
in calculations are: $C=2,\epsilon=1.2,\beta=0.1.$ Also, the
constant of numerical integration of the last term of (31) is
chosen to the radiation force be zero at $y->\infty.$

All pictures show that streaming develop with some delay after the
source passing. The reason is non-local relation between the
source and induced streaming.The velocity field may be calculated
accordingly to the formula (30) completed with equations for the
other components of $\overrightarrow{V} $. At the first stage of
evolution, the nonlinear term $(\overrightarrow
{V}\overrightarrow{\nabla})\overrightarrow{V}$ may be neglected
and the linear equation like this of thermal conductivity with
viscous coefficient cased by shear viscosity $\delta_{1}^{2}$
should be solved. The acoustic source in the right-hand side of
(30) is namely the radiation force. The temporal behavior of the
radiation force at the two longitudinal coordinates $y=1$ and
$y=3$ is presented at the figures 3,4, $x=\sqrt{2}/2.$ The
mono-polar source is naturally attenuated during its propagation
along y-axis. Note that the radiation force is multiplied by the
large value $\frac{1}{\sqrt{\mu}\beta}$ at all pictures, so the
real scale is much less.

%\newpage

\begin{center}
\epsfig{file=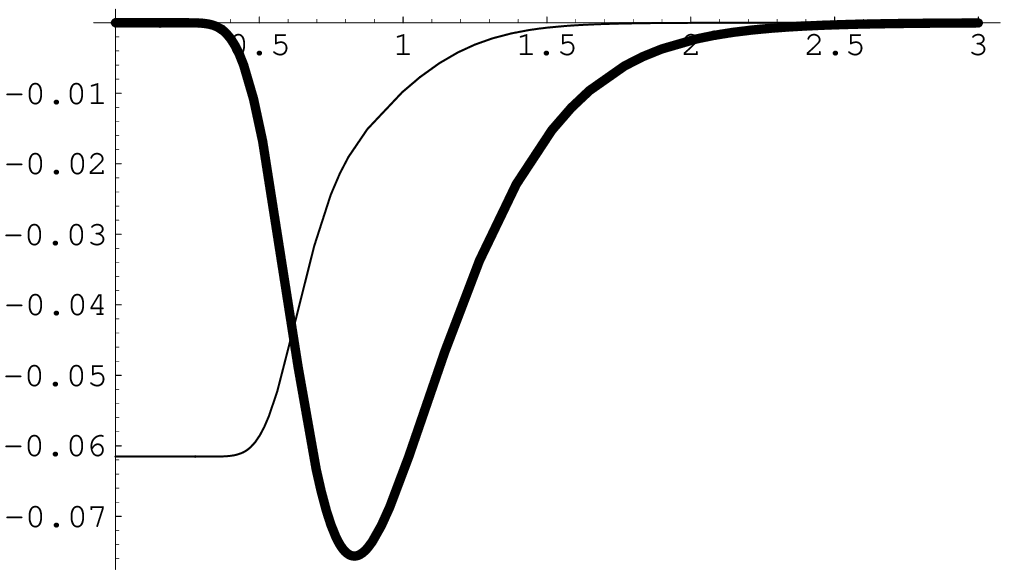, height=4cm, width=8.6cm ,clip=,angle=0}\\
Fig.1 Longitudinal distribution of the radiation force
$\frac{F_{1x}}{\sqrt{\mu }\beta}(y)$ (thin line) and the acoustic
source $\rho_{1}(y) $ (bold line) at $t=1,x=\sqrt{2}/2.$
\end{center}

\begin{center}
\epsfig{file=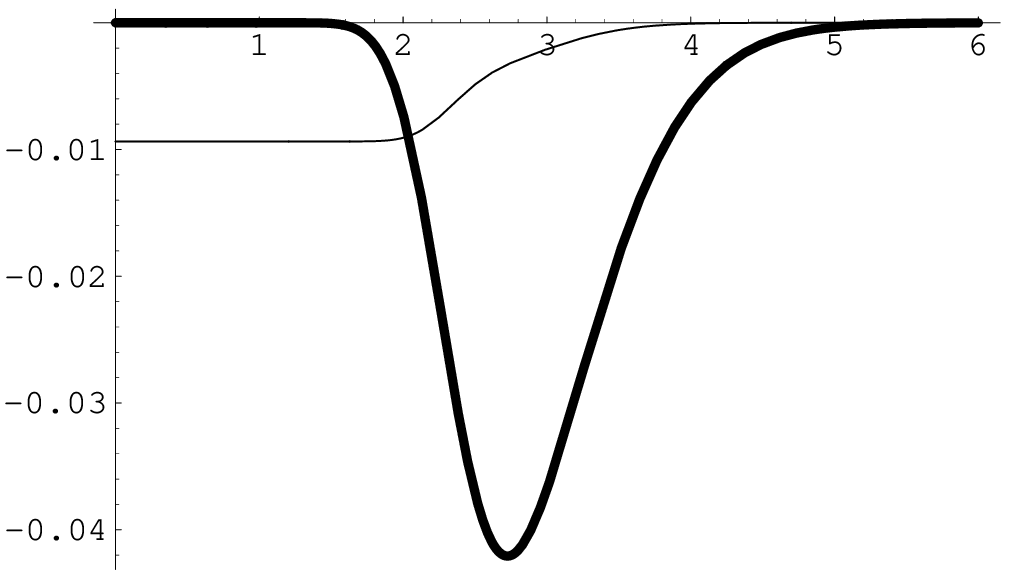, height=4cm, width=8.6cm ,clip=,angle=0}\\
Fig.2 Longitudinal distribution of the radiation force
$\frac{F_{1x}}{\sqrt{\mu }\beta}(y)$ (thin line)
 and
the acoustic source $\rho_{1}(y) $ (bold line) at
$t=3,x=\sqrt{2}/2.$
\end{center}

\begin{center}
\epsfig{file=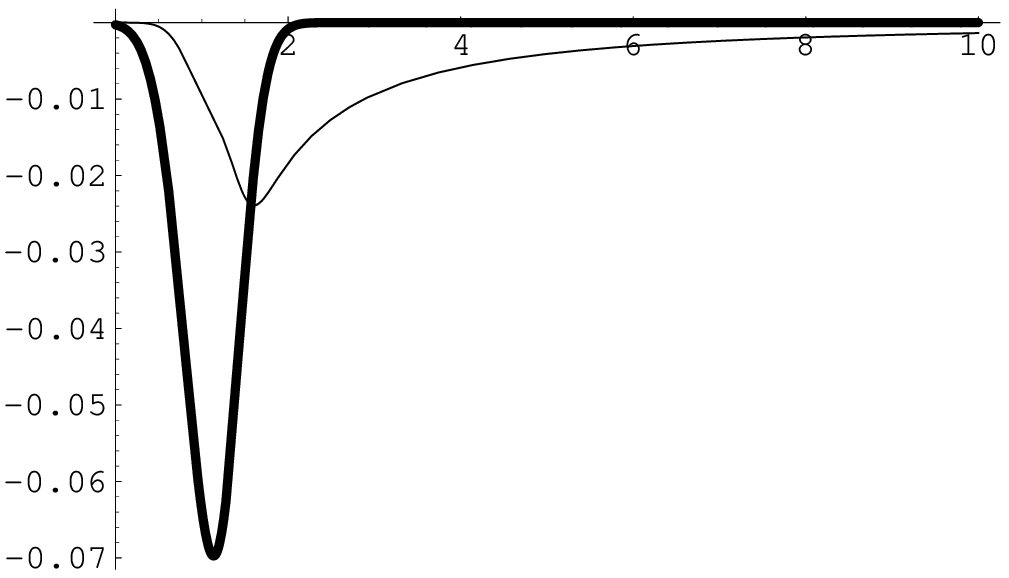, height=4cm, width=8.6cm ,clip=,angle=0}\\
Fig.3 The temporal distribution of the radiation force
$\frac{F_{1x}}{\sqrt{\mu }\beta}(t)$ (thin line)  and
the acoustic source $\rho_{1}(t) $ (bold line) at $y=1,x=\sqrt{2}/2.$%
\end{center}

\begin{center}
\epsfig{file=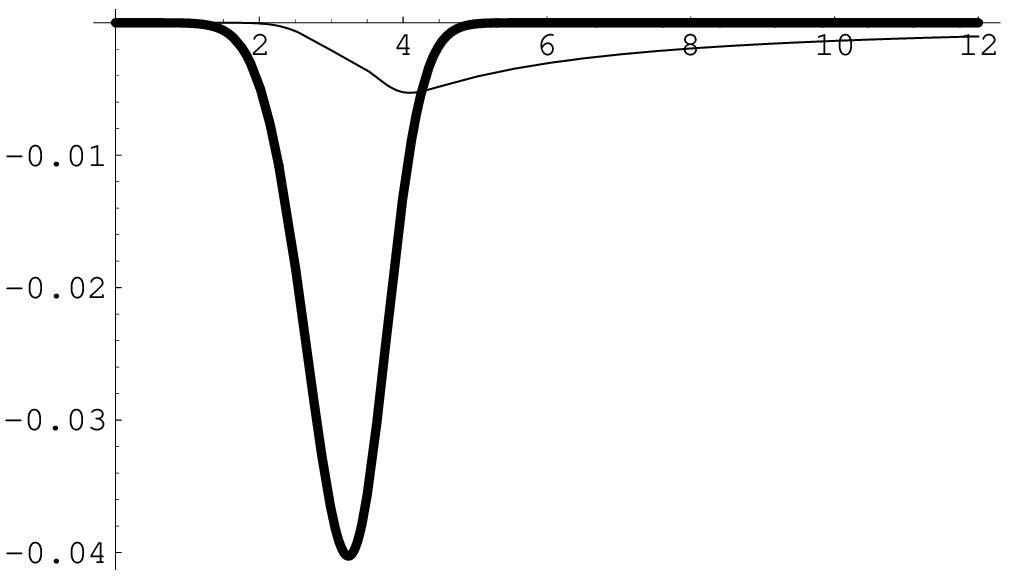, height=4cm, width=8.6cm ,clip=,angle=0}\\
Fig.4 The temporal distribution of the radiation force
$\frac{F_{1x}}{\sqrt{\mu }\beta}(t)$ (thin line) and the acoustic
source $\rho_{1}(t) $ (bold line) at $y=3,x=\sqrt{2}/2.$
\end{center}

%\newpage

\section{Conclusions}

The basic idea of the paper is to separate modes accordingly to their
properties in the weakly nonlinear flow. At first, modes as eigenvectors of
the linear flow should be defined. In the other words, the relations of the
specific perturbations inside every mode should be established. Both
homogeneous and inhomogeneous backgrounds ( see paper on interacting modes in
bubbly liquid [10]), media affected by external forces including the
gravitational one which changes the background density and pressure [11] are
treated in the algorithmic way. The definition of modes is unique, determined
by the linearized differential conservation equations only. Proper initial
or/and boundary conditions are given by the proper superposition of modes.

The second step is to get coupled nonlinear equations for the interacting
modes and to solve it approximately. If one of the acoustic modes is dominant
the generation of the entropy mode and the other acoustic mode in the plane
geometry (namely heating and the reflected wave) is governed by the simple
analytical formulae. The evolution equations may be corrected up to the higher
order nonlinear terms due to increasing influence of the other generated modes [10,12].

The principal advance in comparison to the widely used approaches
consists in\ possibility to treat any sources: initial mixture of
all possible types of motion, non-periodic acoustic source and so
on. In the present paper, radiation force caused streaming is
calculated for any acoustic source, an example of the monopolar
pulse source is considered, the limit of periodic source is
traced.

\protect\bigskip REFERENCES

[1] Zarembo, L.K.(1971). Acoustic streaming. Rozenberg, L.D., editor,
High-intensity ultrasonic fields.Part 3, Plenum Press,
 NewYork,137-191.

[2] Rudenko, O.V., and Soluyan, S.I. (1977). Theoretical
foundations of nonlinear acoustics. Consultants Bureau, New York,
English edition.

[3] Nyborg, W.L. (1997). Acoustic streaming. Hamilton M.F.,
D.T.Blackstock, D.T., editors, Nonlinear Acoustics, Academic
press, New York,$\ $207-231.

[4] Tjotta,S, and Tjotta, J.N. (1993). Acoustic streaming in ultrasound
beams.Proc.13 Int. Symp. Nonlinear Acoustics, 601-606.

[5] Qi,Q.(1993).The effect of compressibility on acoustic streaming near a
rigid boundary for a plane travellimg wave. J. Acoust. Soc.
 Am. 94, 1090-1098.

[6] Menguy,L.,Gilbert, J.(2000). Non-linear acoustic streaming accompanying a
plane stationary wave in a guide, Acta Acustica, 86,
 249-259.

[7] Kamakura,T., Kazuhisa M., Kumamoto. Y., and Breazeale, M.A. (1995).
Acoustic streaming induced in focused Gaussian beams,
 J.Acoust. Soc. Am. 97(5), Pt. 1, 2740-2746.

[8] Makarov,S., and Ochmann, M.(1996). Nonlinear and thermoviscous phenomena
in acoustics, part I., Acustica, 82, 579-606.

[9] Chu, B.-T., and Kovasznay, L.S.G. (1958) Nonlinear interactions in a
viscous heat-conducting compresible gas, Journ. Fluid.
 Mech.,3,$\ $494-514.

[10] Perelomova, A.A.(2000) Projectors in nonlinear evolution
problem: acoustic solitons of bubbly liquid. Applied Mathematical
Letters 13, 93-98.

[11] Perelomova, A.A.(1998) Nonlinear dynamics of vertically propagating
acoustic waves in stratified atmosphere, Acta Acustica, 84 ,
 1002-1006.

[12] Perelomova, A.A. (2001) Directed acoustic beams interacting with heat
mode: coupled nonlinear equations and modified KZK
 equation, Acta Acustica, 87, 176-183.

[13] Leble, S.B. (1991) Nonlinear waves in waveguides with stratification,
Springer-Verlag, Berlin.

[14] Zabolotskaya,E.A, Khokhlov, R.V , Quasiplane waves in the
nonlinear acoustic field of confined beams, Sov.Phys.Acoust.,
 15(1969),35-40

[15] Bakhvalov, N.S.,Zhileikin,Ya.M and Zabolotskaya E.A, Nonlinear theory of
sound beams (American Institute of Physics, New
 York,$\ $1987).

[16]\ Gusev, V.E., and Rudenko, O.V. (1979) Non-steady
quasi-one-dimensional acoustic streaming in unbounded volumes with
 hydrodynamic nonlinearity, Sov.
Phys. Acoust. 25,493-497.

[17] Gusev, VE., and Rudenko, O.V. (1980) Evolution of nonlinear
two-dimensional acoustic streaming in the field of a highly attenuated
 sound beam, Sov. Phys. Acoust. 27(6), 481-484.

[18] Rudenko, O.V., Soluyan S.I. (1977) Theoretical foundations of nonlinear
acoustics, Plenum, New York.
\end{document}